\documentclass[onecolumn,prd,showpacs,preprintnumbers,amsmath,amssymb,floatfix]{revtex4}

\usepackage{graphicx}

\usepackage{bm}
\usepackage{amsfonts}
\usepackage{lineno,hyperref}
\usepackage{array}
\usepackage{microtype}
\usepackage{float}
\usepackage{epstopdf}
\usepackage{mathrsfs}
\usepackage{color}

\begin{document}

\title{Bouncing models in an extended gravity theory}

\author{S. K. Tripathy}
\email{tripathy\_ sunil@rediffmail.com}
\address{Department of Physics, Indira Gandhi Institute of Technology, Sarang, Dhenkanal 759146, Odisha, India}

\author{B. Mishra}
\email{bivudutta@yahoo.com}
\address{Department of Mathematics, Birla Institute of Technology and Science-Pilani, Hyderabad Campus, Hyderabad 500078, India}

\author{Saibal Ray}
\email{saibal@associates.iucaa.in}
\affiliation{Department of Physics, Government College of Engineering and Ceramic Technology, Kolkata 700010, West Bengal, India}

\author{Rikpratik Sengupta}
\email{rikpratik.sengupta@gmail.com}
\address{Department of Physics, Government College of Engineering and Ceramic Technology, Kolkata 700010, West Bengal, India}

\begin{abstract}
Some bouncing models are investigated in the framework of an extended theory of gravity. The extended gravity model is a simple extension of the General Relativity where an additional matter geometry coupling is introduced to account for the late time cosmic speed up phenomena. The dynamics of the models are discussed in the background of a flat FRW universe.  Some viable models are reconstructed for specifically assumed bouncing scale factors. The behavior of the models are found to be decided mostly by the parameters of the respective models. The extended gravity based minimal matter-geometry coupling parameter has a role to remove the omega singularity occurring at the bouncing epoch. It is noted that the constructed models violate the energy conditions, however, in some cases this violation leads to the evolution of the models in phantom phase. The stability of the models are analyzed under linear homogeneous perturbations and it is found that, near the bounce, the models show instability but the perturbations decay out smoothly to provide stable models at late times.


\end{abstract}

\maketitle

\section{Introduction} 
Observations from high redshift supernova, Cosmic Microwave Background (CMB) radiation anisotropy, large scale structure and Baryon Acoustic Oscillation have confirmed that the universe  is accelerating at the present epoch. This late time cosmic speed up issue has led to the development of many new ideas and concept. In the framework of General Relativity(GR), this phenomenon has been attributed to an exotic energy form known as dark energy (DE).  In its usual form, GR is not able to explain this bizarre fact. Therefore, additional dynamical degrees of freedom  mediating through scalar fields are required~\cite{Zlatev1999,Steinhardt1999}. In GR, this DE corresponds to a mysterious fluid with negative pressure. Simplest possible DE candidate is the cosmological constant~\cite{Sahni2008,Weinberg1989}.  However, the cosmological constant is plagued by many fundamental issues such as the fine tunning problem and coincidence problem.  In order to avoid the cosmological constant problem, slowly rolling scalar fields like quintessence have been proposed~\cite{Copeland2006,Caldwell2009}. Most of the dynamical scalar fields are ghost fields with unusual negative kinetic energy at least around flat, cosmological or spherically symmetric backgrounds, e.g. Bulware-Desser mode in massive gravity~\cite{Bulware1972}, bending mode in the self-accelerating branch of Dvali-Gabadadze-Porrati model~\cite{Koyama2007,Sbisa2015,Gumrukcuoglu2016}.  

Geometrically modified theories of gravity have been proposed in recent times as alternatives to GR, that do not require additional degrees of freedom to explain the late time cosmic acceleration and therefore provide ghost free models.  In this alternative approach, the Einstein-Hilbert action is modified by considering  a more general function of $R$ or by a  matter-geometry coupled functional $f(R,T)$ in place of the Ricci scalar $R$. There are so many geometrically modified theories of gravity have been proposed in literature. Notable among them are  $f(R)$ theory~\cite{Caroll2004,Nojiri2007,Bertolami2007}, $f(G)$ gravity~\cite{Nojiri2005,Li2007}, $f(\mathcal{T})$ theory~\cite{Linder2010,Myrzakulov2011} and $f(R,T)$ theory~\cite{Harko2011}. Among these modified theories, the $f(R,T)$ gravity theory~\cite{Harko2011} has attracted a lot of research attention in recent times and has been widely used to address many issues in cosmology and astrophysics~\cite{Zubair2015,Mishra2016,Yousaf2016,Singh2018,Tretyakov2018,Velten2017,Wu2018,Baffou2017,Carvalho2017,Mishra2018a,Baffou2019,Houndjo2014,Das2017,Deb2018,Biswas2020}. Extended gravity theories are a specific extension of the $f(R,T)$ gravity theory where the functional  $f(R,T)$ contains a term  linear in $R$. Recently, there has been an increased interest in extended theories of gravity because of its simple structure and elegance in addressing issues in cosmology~\cite{Capozziello2011,Mishra2018b,Mishra2018c,Capozziello2019,Tripathy2020}.

The cosmological evolution in the early Universe is usually described by the Standard Big Bang cosmology. However, the Standard  cosmology model suffers from many issues such as the flatness problem, the horizon problem, initial singularity and baryon asymmetry problem. The flatness problem and the Horizon problem are concerned with the questions: why is the density of universe so close to the critical density? and   why does the universe (or CMB radiation) look the same in all directions? The inflationary scenario solved some of these issues of the early Universe standard model and provided a causal theory of structure formation~\cite{Guth1981,Mukhanov1981}. In spite of great success, the inflationary scenario suffers from the singularity problem and the trans-Planckian problem for fluctuations. At the onset of inflation, the Universe undergoes an almost exponential expansion. Before such an inflationary phase, as an obvious phenomenon singularity occurs. Therefore inflationary scenario fails to reconstruct the complete past history of the Universe. 

An alternative model to the Standard Big Bang scenario, describing inflation without initial singularity had been proposed which is known as the emergent universe~\cite{Ellis2004}. Another non-singular approach without inflation known as the matter bounce scenario have been proposed~\cite{Bars2011,Bars2012,Brandenberger2011,Brandenberger2012}. For some reviews see~\cite{Novello2008,Brandenberger2016,Battefeld2014}. 

The initial singularity occurring in the Standard Big Bang cosmology and the inflationary cosmology can be suitably avoided in the matter bounce scenario. In bouncing scenario, the Universe undergoes an initial matter dominated contraction phase followed by a non-singular bounce and then there is a causal generation for fluctuation~\cite{Novello2008,Brandenberger2016}. Moreover, in the contracting phase the scale factor decreases ($\dot{a}<0$) and in the expanding phase, scale factor increases ($\dot{a}>0$) and at the matter bounce epoch, we have $\dot{a}=0$. Consequently, in bouncing cosmology, the Hubble parameter $H$ passes across zero ($H=0$) from  negative values $H<0$ to  positive values $H>0$~\cite{Tripathy2019,Tripathy2020a}.  Bouncing cosmologies have been investigated in alternative gravity theories such as $f(R)$ theory~\cite{Bamba2014a,Barragan2009,Barragan2010,Saidov2010,Chakraborty2018}, modified Gauss-Bonnet gravity~\cite{Bamba2014b,Bamba2015}, $f(R,T)$ gravity~\cite{Singh2018,Tripathy2019,Sahoo2020a} and $f(T)$ gravity~\cite{Cai2011}. Resolving the initial singularity problem applying Loop Quantum Gravity approach gives rise to a cosmological scenario which can be considered to be a combination of the emergent and bouncing universe picture~\cite{Alesci2017}.

In  the present work, we have investigated some bouncing models in the framework of an extended gravity theory. Our motivation is to use a simple extended gravity theory exploring the  geometrical degrees of freedom to explain the late time cosmic speed up phenomenon and to investigate the bouncing behavior at an initial epoch. We have considered some widely known bouncing models described through specific scale factors in the framework of extended gravity and studied the cosmic dynamics. The paper is organized as follows: in Sec. II, the basic formalism of the extended gravity theory and the field equations for a flat FRW universe have been derived. In Sec. III, dynamical physical parameters such as the energy conditions and the equation of state (EOS) parameter are derived in terms of the Hubble parameter. Three different bouncing models along with their evolutionary behavior have been studied in Sec. IV. At the end the conclusion and summary are presented in Sec. V.

\section{Basic Formalism of the extended gravity}
The action for a geometrically modified extended theory with a matter-geometry coupling is considered as 
\begin{equation} \label{eq:1}
S=\int d^4x\sqrt{-g}\left[\frac{1}{2} \left(R+\beta f(T)\right)+ \mathcal{L}_m \right],
\end{equation}
where $\mathcal{L}_m$ is the matter Lagrangian. $f(T)$ is an arbitrary function of the trace $T$ of the energy-momentum tensor. The unit system followed in writing the action is the natural unit system where $8\pi G=c=1$; $G$ and $c$ being the Newtonian gravitational constant and speed of light in vacuum respectively. The interesting fact of the above action is that it reduces to that of GR for a vanishing $\beta$. 

Varying this action with respect to the metric $g_{\mu\nu}$, the modified field equations are obtained as
\begin{equation} \label{eq:2}
R_{\mu\nu}-\frac{1}{2}Rg_{\mu\nu}=\left[1 -\beta f_{T}(T)\right]T_{\mu\nu}+ \frac{1}{2}\beta f(T)g_{\mu\nu}  -\beta f_T(T)\Theta_{\mu\nu},
\end{equation}
where the energy-momentum tensor $T_{\mu\nu}$ is defined through the matter Lagrangian as
\begin{equation}\label{eq:3}
T_{\mu\nu}=-\frac{2}{\sqrt{-g}}\frac{\delta\left(\sqrt{-g}\mathcal{L}_m\right)}{\delta g^{\mu\nu}},
\end{equation}
and $\Theta_{\mu\nu}=g^{\rho\sigma}\frac{\delta T_{\rho\sigma}}{\delta g^{\mu\nu}}$. Here $f_{T}(T)$ denotes the partial derivative of $f(T)$ with respect to $T$. 

We chose a perfect fluid distribution of the universe with energy density $\rho$ and pressure $p$ for which the energy-momentum tensor is given by
\begin{equation}\label{eq:4}
T_{\mu\nu}=-pg_{\mu\nu}+\left(\rho+p\right)u_{\mu}u_{\nu},
\end{equation}
where $u_{\mu}$ is the time like four-velocity vector of the cosmic fluid that satisfies the relation $u_{\mu}u^{\mu}=1$ in co-moving coordinates. In this context we also chose the matter Lagrangian as $\mathcal{L}_m=-p$, so that $\Theta_{\mu\nu}=-pg_{\mu\nu}-2T_{\mu\nu}$. 

The modified field equation now reduces to
\begin{equation} \label{eq:5}
R_{\mu\nu}-\frac{1}{2}Rg_{\mu\nu}=\left[1 +\beta f_{T}(T)\right]T_{\mu\nu}+\left[f_{T}(T)p+\frac{1}{2}f(T)\right]\beta g_{\mu\nu}.\
\end{equation}

Simple algebraic manipulation leads to 
\begin{equation}\label{eq:6}
G_{\mu\nu}= \kappa_{T}\left[T_{\mu\nu}+ T^{int}_{\mu\nu}\right],
\end{equation}
where $G_{\mu\nu}= R_{\mu\nu}-\frac{1}{2}Rg_{\mu\nu}$ is the usual Einstein tensor and $\kappa_{T}= 1 +\beta f_{T}(T)$ is the redefined Einstein constant. The redefined Einstein constant may be a dynamical quantity for a non linear choice of the functional $f_{T}(T)$. However, for a linear choice of the functional $f_{T}(T)$ it becomes a constant carrying the signature of the extended gravity through the coupling parameter $\beta$.  In Eq. \eqref{eq:6}, we have 
\begin{equation}\label{eq:7}
T^{int}_{\mu\nu}=\frac{1}{\kappa_T}\left[ f_{T}(T)p+\frac{1}{2}f(T)\right]\beta g_{\mu\nu}.
\end{equation}

The geometrical modification in the action leads to the extra effective interaction term $T^{int}_{\mu\nu}$ which is necessary to provide the required acceleration at late times. It is worthy to mention here that, such a coupling of matter and curvature is motivated from quantum effects and particle creation process. In the limit $\beta \rightarrow 0$, the interaction term vanishes. In principle one may construct viable cosmological models through suitable choices of the functional $f_2(T)$  which may be confronted with recent observations. In the present work, we consider a linear functional
\begin{equation}\label{eq:8}
\frac{1}{2}f(T)= T,
\end{equation}
so that
\begin{eqnarray}
\kappa_T = 1+2\beta,\label{eq:9}
\end{eqnarray}
and
\begin{equation}
T^{int}_{\mu\nu} = \frac{1}{\kappa_T}\left(2p+T\right) \beta g_{\mu\nu}.\label{eq:10}
\end{equation}

The field equations in the extended gravity theory for a flat FRW model become
\begin{eqnarray}
3H^2 &=& \alpha \rho-\beta p,\label{eq:11}\\
2\dot{H}+3H^2 &=& -\alpha p+\beta \rho.\label{eq:12}
\end{eqnarray}

Here $\alpha=1+3\beta$ and we denote the ordinary time derivatives as overhead dots. The Hubble parameter is defined through the scale factor $a(t)$ as $H=\frac{\dot{a}}{a}$.

\section{Physical parameters}
The physical parameters of the models such as the pressure and energy density  can be obtained from the field equations \eqref{eq:11} and \eqref{eq:12} in terms of the Hubble parameter as
\begin{eqnarray}
p &=& -\frac{3\kappa_T H^2+2\alpha\dot{H}}{\alpha^2-\beta^2},\label{eq:13}\\
\rho &=& \frac{3\kappa_T H^2-2\beta\dot{H}}{\alpha^2-\beta^2}.\label{eq:14}
\end{eqnarray}

The energy condition $\rho+p\geq 0$ and $\rho\geq 0$ put extra constraints on the model. The positive energy density constrains the coupling parameter through the condition $\frac{3\kappa_T}{2\beta}\geq \frac{\dot{H}}{H^2}$.

The equation of state (EOS) parameter can be obtained as
\begin{equation}
\omega = -\frac{3\kappa_T H^2+2\alpha\dot{H}}{3\kappa_T H^2-2\beta\dot{H}}.\label{eq:15}
\end{equation}

The EOS parameter can also be expressed in terms of the deceleration parameter $q=-1-\frac{\dot{H}}{H^2}$ as
\begin{equation}
\omega=-1+\frac{2(\alpha+\beta)(1+q)}{3\kappa_T+2\beta(1+q)}.\label{eq:16}
\end{equation}

This is an important equation in the sense that, the time dependence of the deceleration parameter controls the time evolution of the equation of state parameter. A constant EOS will result from a constant deceleration parameter.

The null energy condition (NEC) is expressed as $\rho+p\geq 0$. In the purview of the extended gravity theory, from Eqs. \ref{eq:13} and \ref{eq:14}, we can have 
\begin{equation}
\rho+p=-\frac{2}{\kappa_T}\dot{H}.\label{eq:17}
\end{equation}

Since $\kappa_T=1+2\beta$ is a positive quantity for a small value of $\beta$, the NEC is violated for positive values of the Hubble parameter. In fact, in phantom models, the NEC is usually violated.

\section{Bouncing cosmologies}
In this section we have considered three different scale factors that show bouncing behavior at some epochs. The bouncing nature of these scale factors are discussed.  Also, we derive the dynamical properties such as the deceleration parameter and the EOS parameter for these models. 

\subsection{Model I}
Let us consider a bouncing scale factor~\cite{Bamba2014b}
\begin{equation}\label{eq:18}
a(t)=a_1e^{\lambda_1 t}+a_2 e^{-\lambda_2 t}, 
\end{equation}
where $a_1, a_2, \lambda_1$ and $\lambda_2$ are positive constants. Also we assume that $a_2\neq 0$ and $\lambda_2 \neq 0$. For $a_2=0$ or $\lambda_2 = 0$, a de Sitter universe is obtained.

From the scale factor, we can have
\begin{equation}\label{eq:19}
\dot{a} = a_1 \lambda_1e^{\lambda_1 t}-a_2 \lambda_2e^{-\lambda_2 t}.
\end{equation}

At the bouncing epoch, $\dot{a}=0$. This condition decides the time of the bouncing scenario as
\begin{equation}\label{eq:20}
t_b=\frac{1}{\lambda_1+\lambda_2}ln\left(\frac{a_2\lambda_2}{a_1\lambda_1}\right).
\end{equation}

For $\lambda_1=\lambda_2=\lambda$, we have $t_b=ln\left(\frac{a_2}{a_1}\right)^{\frac{1}{2\lambda}}$. If we wish that the bounce should  occur at $t_b=0$, then the parameter should satisfy the relation $a_1\lambda_1=a_2\lambda_2$ in general and $a_1=a_2$ for $\lambda_1=\lambda_2=\lambda$.

The Hubble parameter for the bouncing scale factor in Eq. \eqref{eq:18} is given by
\begin{equation}
H=\frac{\dot{a}}{a}=\frac{a_1\lambda_1e^{\lambda_1t}-a_2\lambda_2e^{-\lambda_2t}}{a_1e^{\lambda_1t}+a_2e^{-\lambda_2t}}, \label{eq:21}
\end{equation}

and consequently 
\begin{equation}
\dot{H}=\frac{a_1\lambda_1^2e^{\lambda_1t}+a_2\lambda_2^2e^{-\lambda_2t}}{a_1e^{\lambda_1t}+a_2e^{-\lambda_2t}}-\left[\frac{a_1\lambda_1e^{\lambda_1t}-a_2\lambda_2e^{-\lambda_2t}}{a_1e^{\lambda_1t}+a_2e^{-\lambda_2t}}\right]^2. \label{eq:22}
\end{equation}

At the bouncing epoch $(t_b=0)$, obviously we have $H=0$ and $\dot{H}=0$. Therefore, the scale factor in Eq. \eqref{eq:18} satisfies all bouncing conditions.

The deceleration parameter for the bouncing scale factor becomes
\begin{equation}
q=-\frac{\left[a_1e^{\lambda_1 t}+a_2 e^{-\lambda_2 t}\right]\left[a_1\lambda_1^2e^{\lambda_1t}+a_2\lambda_2^2e^{-\lambda_2t}\right]}{\left[a_1\lambda_1e^{\lambda_1t}-a_2\lambda_2e^{-\lambda_2t}\right]^2}. \label{eq:23}
\end{equation}

If we consider that $\lambda_1=\lambda_2=\lambda$, then we have
\begin{eqnarray}
H &=& \lambda\left[\frac{a_1 e^{\lambda t}-a_2 e^{-\lambda t}}{a_1e^{\lambda t}+a_2e^{-\lambda t}}\right]=\lambda \tanh{\lambda t}. \label{eq:24}\\
\dot{H} &=& \lambda^2- H^2,\label{eq:25}\\
q &=& -\left(\frac{\lambda}{H}\right)^2.\label{eq:26}
\end{eqnarray}

One can note that as $t\rightarrow +\infty$, $H\rightarrow \lambda$  and as $t\rightarrow -\infty$, $H\rightarrow -\lambda$. Also for any cosmic time $t$ before and after the bounce, a positive slope of the Hubble parameter $(\dot{H}>0)$ can be ensured for the condition $\lambda^2 > H^2$. In FIG. 1, the time evolution of the deceleration parameter for the present model is shown as a function of the cosmic time. In order to plot the figure, we have considered $\lambda=2.2$. For the scale factor chosen in Eq. \eqref{eq:18}, the deceleration parameter evolves with time both in the pre and post bounce epochs from a negative value less that $-1$ to an asymptotic value of  $-1$ at large cosmic time. However, at the bouncing epoch, $H=0$ and we have a singularity for the deceleration parameter.

\begin{figure}[ht!]
\centering
\includegraphics[width=8cm,height=6cm,angle=0]{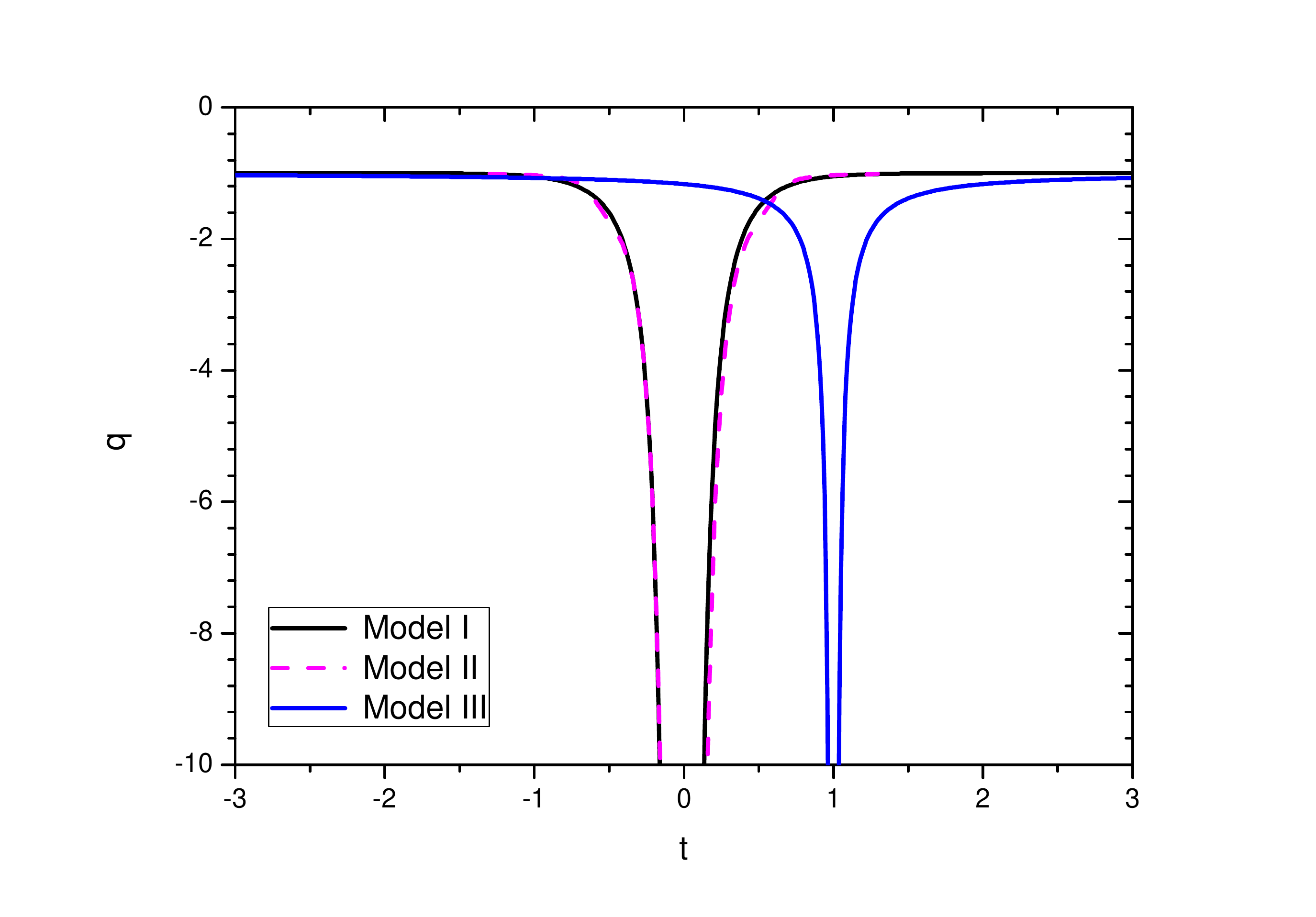}
\caption{Plot for the variation of the deceleration parameter versus time for the three different bouncing models considered in the present work. For Model I and Model II, we chose $\lambda=2.2$. For the Model III, we consider $a_0=1$, $t_0=1$ and $n=0.6$.}
\end{figure}

The energy density and pressure for this bouncing model are obtained as
\begin{eqnarray}
\rho &=& \frac{\left(3\kappa_T+2\beta\right)H^2-2\beta\lambda^2}{\alpha^2-\beta^2},\label{eq:27}\\
p    &=&  -\frac{\left(3\kappa_T-2\alpha\right)H^2+2\alpha\lambda^2}{\alpha^2-\beta^2}.\label{eq:28}
\end{eqnarray}

Since at  bounce, the scale factor obeys the condition $H=0$, we have the energy density at bounce as $\rho=-\frac{2\beta\lambda^2}{\alpha^2-\beta^2}$. In order to ensure positive energy density during the bounce, the minimal matter geometry coupling parameter $\beta$ should be constrained to be a negative quantity. Further, $\beta$ should satisfy the condition, $6\beta > -(1+8\beta^2)$. The evolutionary aspect of the energy density for the present model is shown in FIG. 2 for three representative values of the parameter $\beta$, namely $\beta$= -0.1, -0.2 and -0.24. For these values of $\beta$, the energy density becomes a positive quantity in the positive and negative time domain. If we decrease $\beta$ further beyond $-0.25$, the positive condition of the energy density is not satisfied. The energy density shows a well near the bounce for all the chosen values of $\beta$. However, with an increase in $\beta$, the depth of the well increases. Similarly, in FIG. 3, the pressure is shown for the same three representative values of $\beta$ for the Model I. The pressure is observed to be a negative quantity in both the time domain. Near the bouncing epoch, the curves of the pressure show a bump whose height decreases with an increase in the value of $\beta$. 

\begin{figure}[ht!]
\centering
\includegraphics[width=8cm,height=6cm,angle=0]{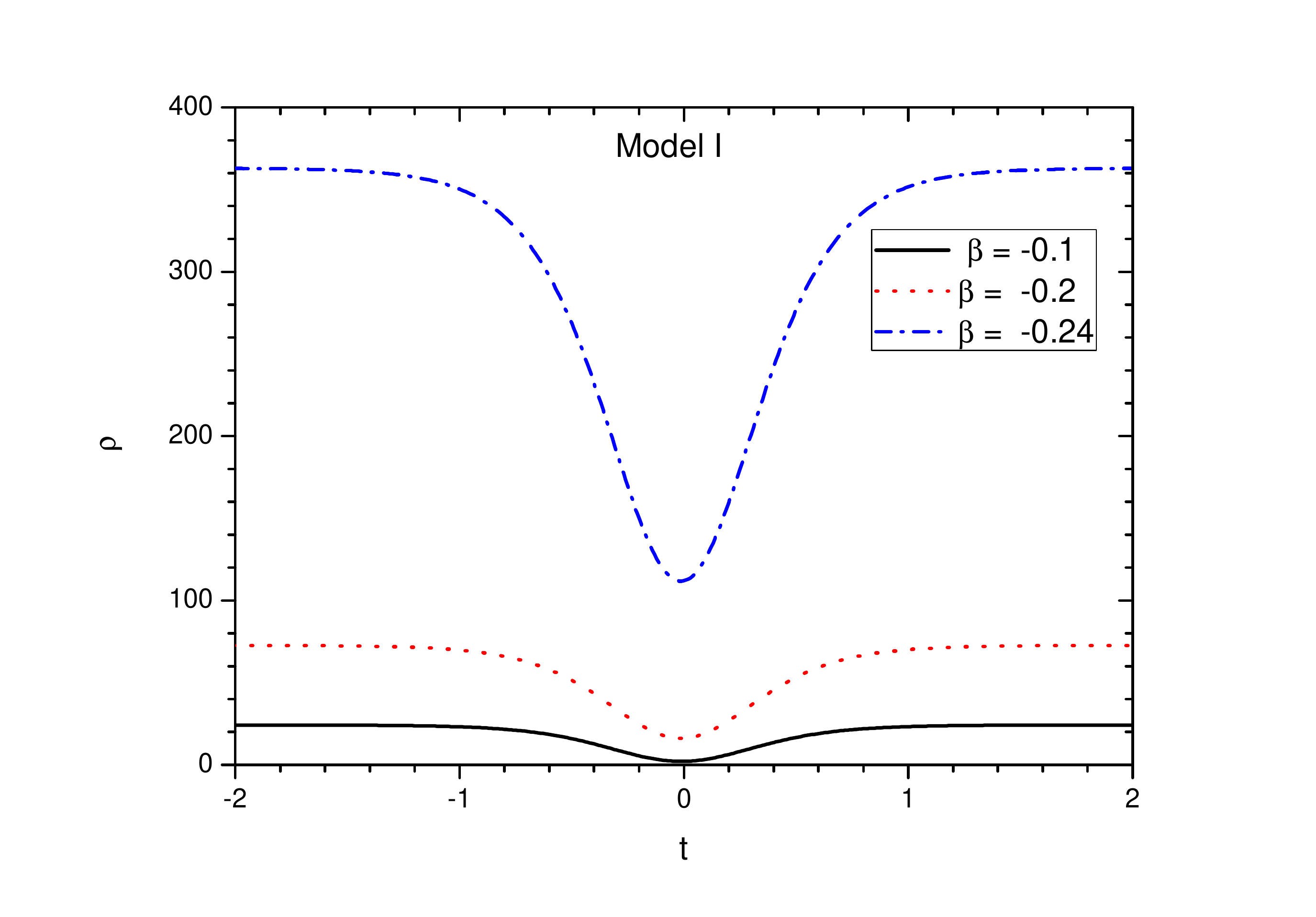} 
\caption{Plot for the variation of the $\rho$ versus $t$ with the specifications for different parametric values (Model I)}
\end{figure}

\begin{figure}[ht!]
\centering
\includegraphics[width=8cm,height=6cm,angle=0]{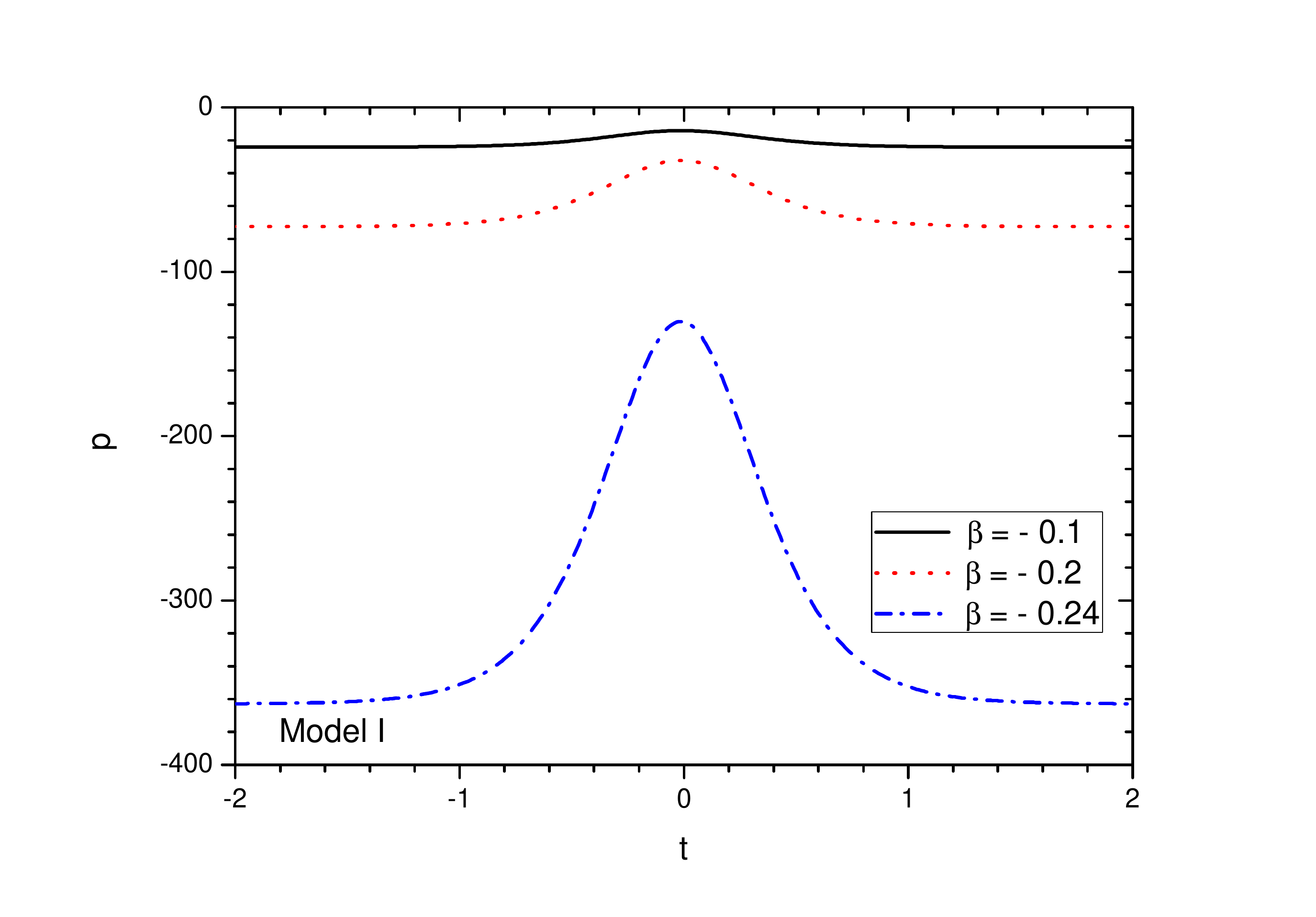}
\caption{Plot for the variation of the $p$ versus $t$ with the specifications for different parametric values (Model I)}
\end{figure}

The EOS parameter for this bouncing model become
\begin{equation}
\omega=-1+\frac{2(1+4\beta)(H^2-\lambda^2)}{(3\kappa_T+2\beta)H^2-2\beta\lambda^2}. \label{eq:29}
\end{equation} 

It is interesting to note that, an evolving deceleration parameter in the present model leads to an evolving EOS parameter in the phantom region. At the bounce, $\dot{H}=0$ and consequently we have
\begin{equation} 
\omega=3+\frac{1}{\beta}.\label{eq:30}
\end{equation}

In GR limit, we have $\beta\rightarrow 0$ leading to a $\omega$-singularity. However, for a finite non-zero value of $\beta$, the $\omega$-singularity is removed. In other words, the present framework of extended gravity with finite values of the minimal matter-geometry coupling parameter $\beta$, lifts the $\omega$-singularity occurring in GR. In FIG. 4, the evolutionary aspect of the EOS parameter for the bouncing model (\eqref{eq:29}) is shown here for different values of $\beta$= -0.1, -0.2 and -0.24.  The EOS parameter evolves to form a well near-bounce for lower values of $\beta$. The depth of the well depends on the choice of the coupling constant $\beta$. We have considered $\lambda=2.2$ for plotting the figures. In general, in both sides of the bouncing epoch, the EOS parameter evolves to overlap with a cosmological constant $\omega=-1$. As we move away from the bouncing epoch, the EOS parameter is observed to evolve in the phantom region and overlaps with a cosmological constant at large time. This behavior of the EOS parameter occurs in both the positive and negative time zone away from the bouncing epoch.

\begin{figure}[ht!]
\centering
\includegraphics[width=8cm,height=6cm,angle=0]{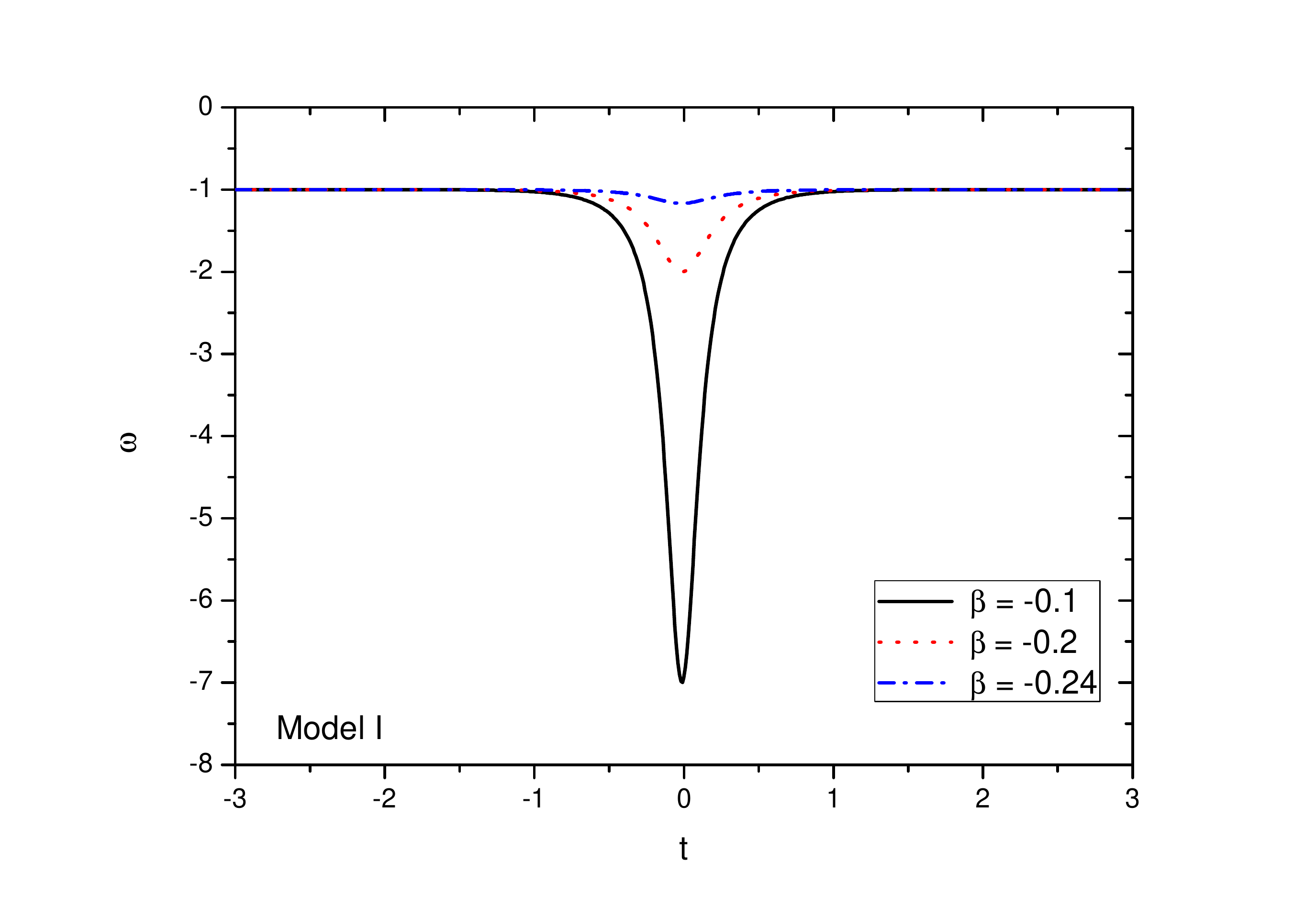} 
\caption{Plot for the variation of the EOS parameter versus time with the specifications of different $\beta$ (Model I)}
\end{figure}

For the bouncing scale factor given in Eq. \eqref{eq:17}, the NEC is given by
\begin{equation}
\rho+p=-\frac{2}{\kappa_T}\left[\frac{a_1\lambda_1^2e^{\lambda_1t}+a_2\lambda_2^2e^{-\lambda_2t}}{a_1e^{\lambda_1t}+a_2e^{-\lambda_2t}}-\left(\frac{a_1\lambda_1e^{\lambda_1t}-a_2\lambda_2e^{-\lambda_2t}}{a_1e^{\lambda_1t}+a_2e^{-\lambda_2t}}\right)^2\right].\label{eq:31}
\end{equation} 

In FIG. 5, the NEC for the present bouncing model is shown for three different values of $\beta$. Since $\dot{H}$ remains positive through out the cosmic evolution before and after the bounce, we have $\rho+p <0$ and therefore the energy condition $\rho+p \geq 0$ is violated in the present model.

\begin{figure}[ht!]
\centering
\includegraphics[width=8cm,height=6cm,angle=0]{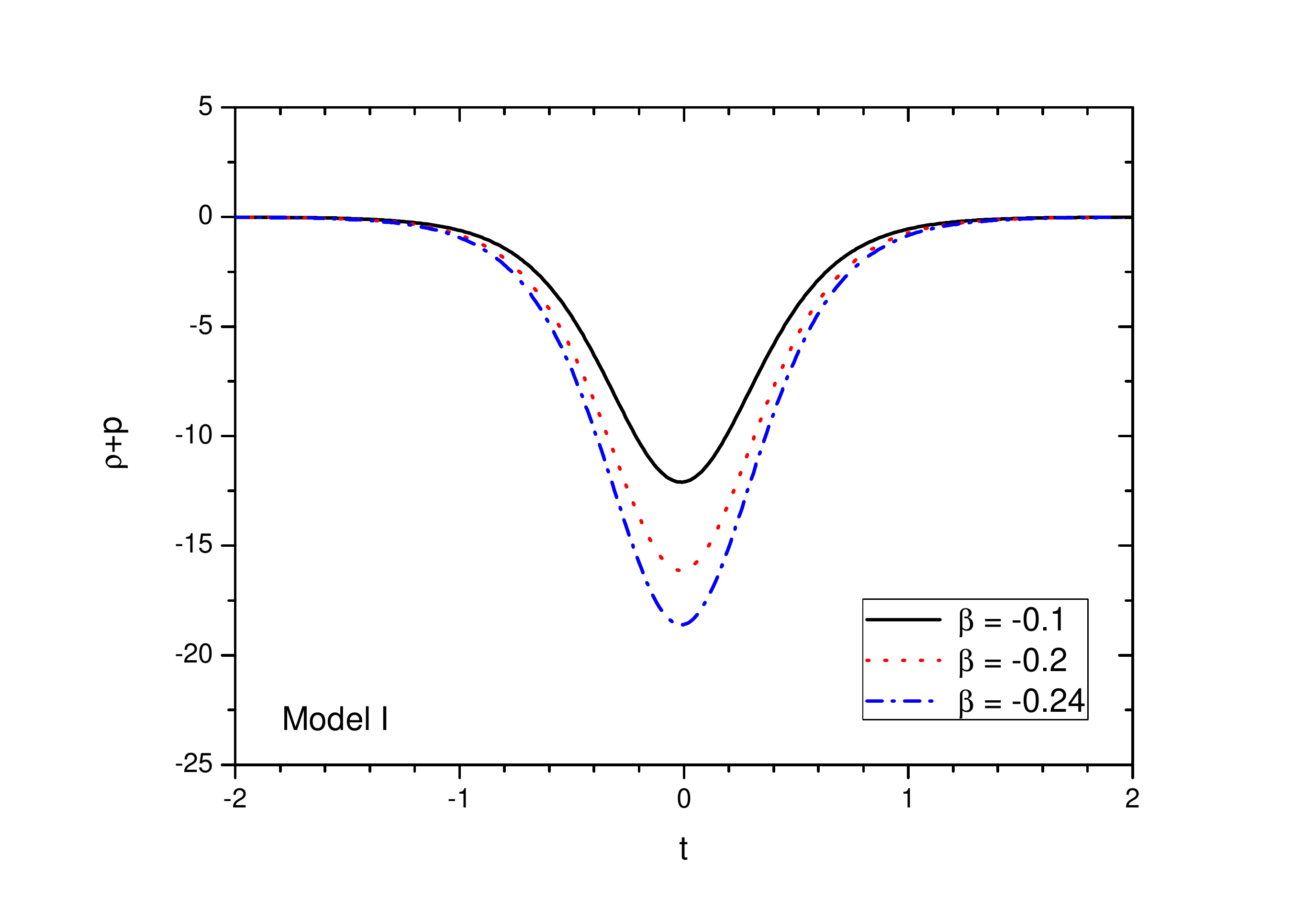}
\caption{Plot for the variation of the $\rho+p$ versus $t$ with the specifications for different parametric values (Model I)}
\end{figure}

\subsection{Model II}
In this case we consider a sum of exponential models  for a bouncing scenario described by a the scale factor~\cite{Bamba2014a}
\begin{equation}
a(t)=\frac{1}{2}\left[e^Y+e^{Y^2}\right],\label{eq:32}
\end{equation}
where $Y=(\lambda t)^2$ and $\lambda$ is a positive constant. One can note that in the limit of $\lambda \rightarrow 0$, the model reduces to the exponential one with the scale factor $a(t)=e^{\lambda^2 t^2}$. In the limit of large $\lambda$, the second term i.e  $e^{Y^2}$ is dominant which may explain a late time cosmic acceleration with positive acceleration $\ddot{a}$.

For the present case, we have
\begin{equation}
\dot{a}=\frac{1}{2}\left[2\lambda^2t e^Y+4\lambda^4t^3e^{Y^2}\right].\label{eq:33}
\end{equation}

Since at $t=0$, we have $\dot{a}=0$, obviously the bounce occurs at $t_b=0$. The Hubble parameter can be expressed as
\begin{equation}
H(t)=2\lambda^2t\left[1+\frac{2\lambda^2t^2-1}{e^{(Y-Y^2)}+1}\right].\label{eq:34}
\end{equation}

Since in the limit $\lambda \ll 1$ we have $e^{Y-Y^2}+1 \simeq 2-Y$. Consequently the Hubble parameter reduces to
\begin{equation}
H \simeq \lambda^2 t.\label{eq:35}
\end{equation}

The slope of the Hubble parameter becomes
\begin{equation}
\dot{H}=2\lambda^2\left(1+2\lambda^2t^2\right)+\frac{16\lambda^8t^6+8\lambda^4t^2-2\lambda^2}{e^{(Y-Y^2)}+1}-H^2,\label{eq:36}
\end{equation}
which reduces to $\dot{H} \simeq \lambda^2 -H^2$ in the limit of $\lambda \ll 1$. In other words, this model behaves as that of the previous model in limit of small exponent $\lambda$. In view of the above, it can be inferred that, this model bounces at $t_b=0$ and has an accelerating behaviour at late times.

The deceleration parameter for this model can be obtained as
\begin{equation}
q=-\left[\frac{(e^{(Y-Y^2)}+1)(1+2\lambda^2t)+(8\lambda^6t^6+4\lambda^2t^2-1)}{t\left(e^{(Y-Y^2)}+1\right)}\right],\label{eq:37}
\end{equation}
which remains negative through out the cosmic evolution for $\lambda \ll 1$ in the positive time domain and remains positive for negative time domain. The behaviour of the deceleration parameter for the present model is shown in FIG. 1. Here we chose the value as $\lambda=2.2$. It is observed that the deceleration parameter evolves from large negative values near the bouncing epoch to become $q \simeq -1$ at late times. As expected, at the bouncing epoch, there occurs a singularity in $q$ because of the vanishing nature of the Hubble parameter at this point of time. It is interesting to note from the figure that, the evolutionary nature of the deceleration parameter for both the Models I and II are similar. 

In FIGS. 6 and 7, the energy density and the pressure for the bouncing model II are shown for the three representative values of $\beta$. While the energy density remains positive through out for all the values of $\beta$, the pressure becomes negative. With an increase in $\beta$, at a given epoch, the energy density decreases but the pressure increases.

\begin{figure}[ht!]
\centering
\includegraphics[width=8cm,height=6cm,angle=0]{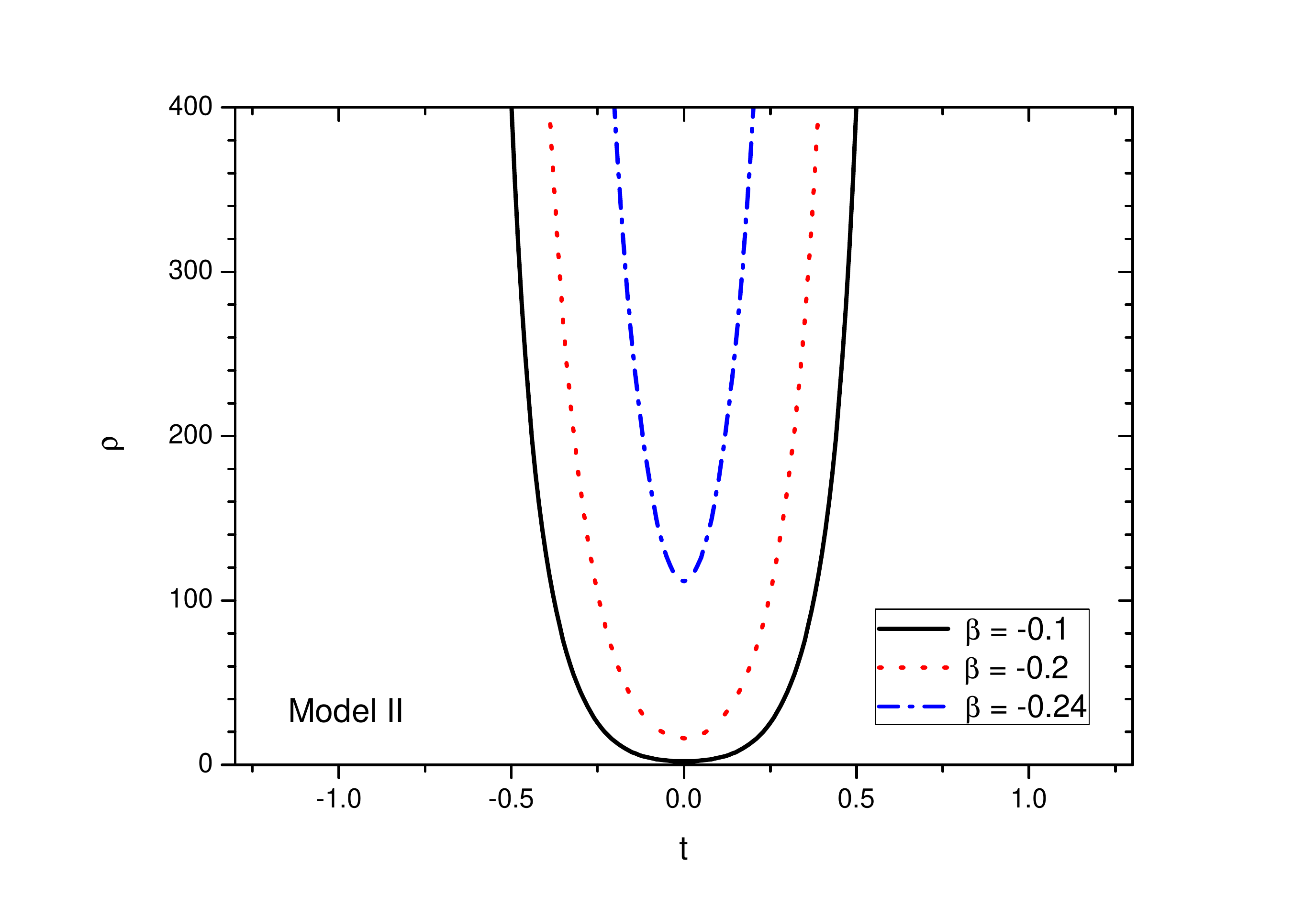}
\caption{Plot for the variation of the $\rho$ versus $t$ with the specifications for different parametric values (Model II)}
\end{figure}

\begin{figure}[ht!]
\centering
\includegraphics[width=8cm,height=6cm,angle=0]{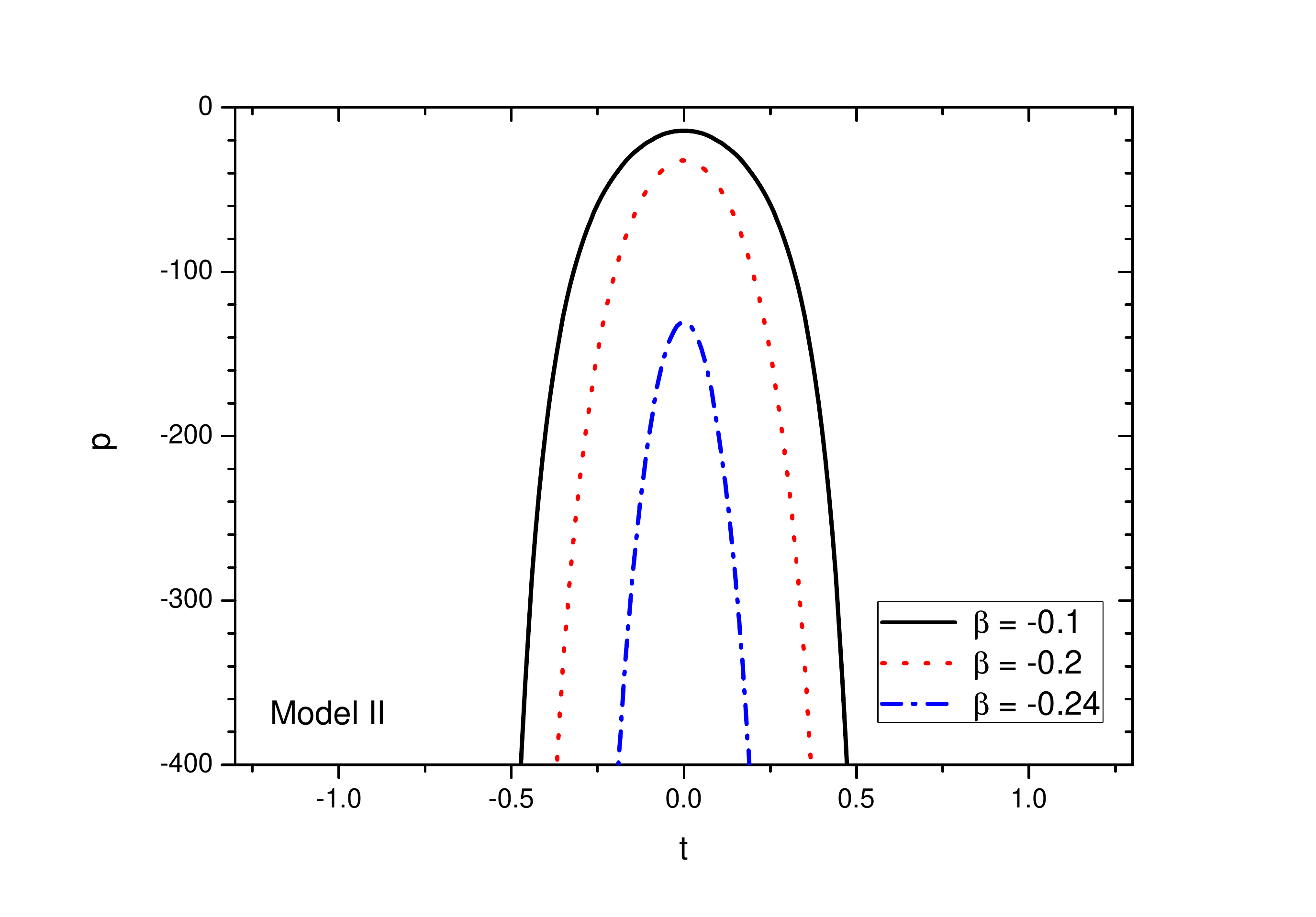}
\caption{Plot for the variation of the $p$ versus $t$ with the specifications for different parametric values (Model II)}
\end{figure}

 From the behaviour of the deceleration parameter, the evolutionary nature of the EOS parameter can be assessed as
\begin{equation} 
\omega= -1+\left[\frac{2(1+4\beta)}{2\beta+\frac{3\kappa_T}{1+q}}\right],\label{eq:38}
\end{equation}
where
\begin{equation}
q = -1 -\left[\frac{(e^{(Y-Y^2)}+1)(2\lambda^2-1)t+1)+(8\lambda^6t^6+4\lambda^2t^2-1)}{t\left(e^{(Y-Y^2)}+1\right)}\right].\label{eq:39}
\end{equation}

For positive time domain, the EOS parameter evolves in the phantom domain. In the limit of large cosmic time, the EOS parameter 
evolves asymptotically to become $\omega \simeq -1$ and the model overlaps with a cosmological constant. The behaviour of the EOS parameter in Eq. (\ref{eq:38}) around the bouncing epoch is shown for the three representative values of $\beta$ in FIG. 8. The value of $\lambda$ is chosen to be 2.2. The behaviour is similar to that of Model I except the fact that, in Model I we have a sharp decrease in $\omega$ compared to that in Model II.

\begin{figure}[ht!]
\centering
\includegraphics[width=8cm,height=6cm,angle=0]{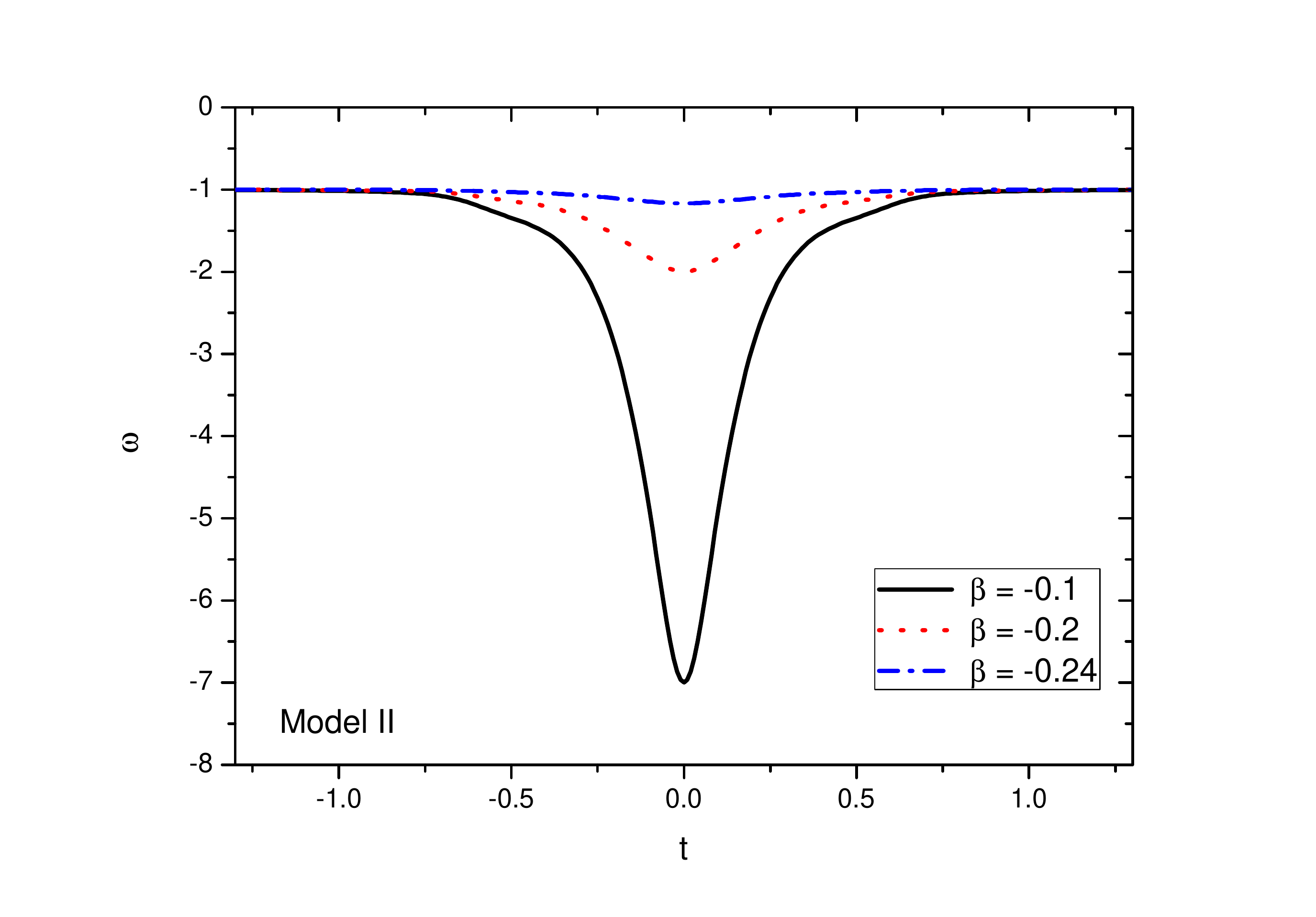} 
\caption{Plot for the variation of the EOS parameter versus time with the specifications of different $\beta$ (Model II)}
\end{figure}

Interestingly, as in the previous model, here also we have at bounce
\begin{equation}
\omega=3+\frac{1}{\beta}.\label{eq:40}
\end{equation}
and  as such signifies the role of the modified gravity in removing the $\omega$-singularity for a finite value of $\beta$.

The present bouncing model with the scale factor as provided in Eq. \ref{eq:32}, the NEC can be obtained from Eq. \ref{eq:17}, as 
\begin{equation}
\rho+p=-\frac{4}{\kappa_T}\left[\lambda^2\left(1+2\lambda^2t^2\right)+\frac{8\lambda^8t^6+4\lambda^4t^2-\lambda^2}{e^{(Y-Y^2)}+1}\right]+\frac{2}{\kappa_T}H^2.
\end{equation}

The NEC for the present model is shown in FIG. 9 for the specified values of $\beta$. As has been stated earlier, it is observed from the figure that for small values of $\beta$, the NEC is violated.

\begin{figure}[ht!]
\centering
\includegraphics[width=8cm,height=6cm,angle=0]{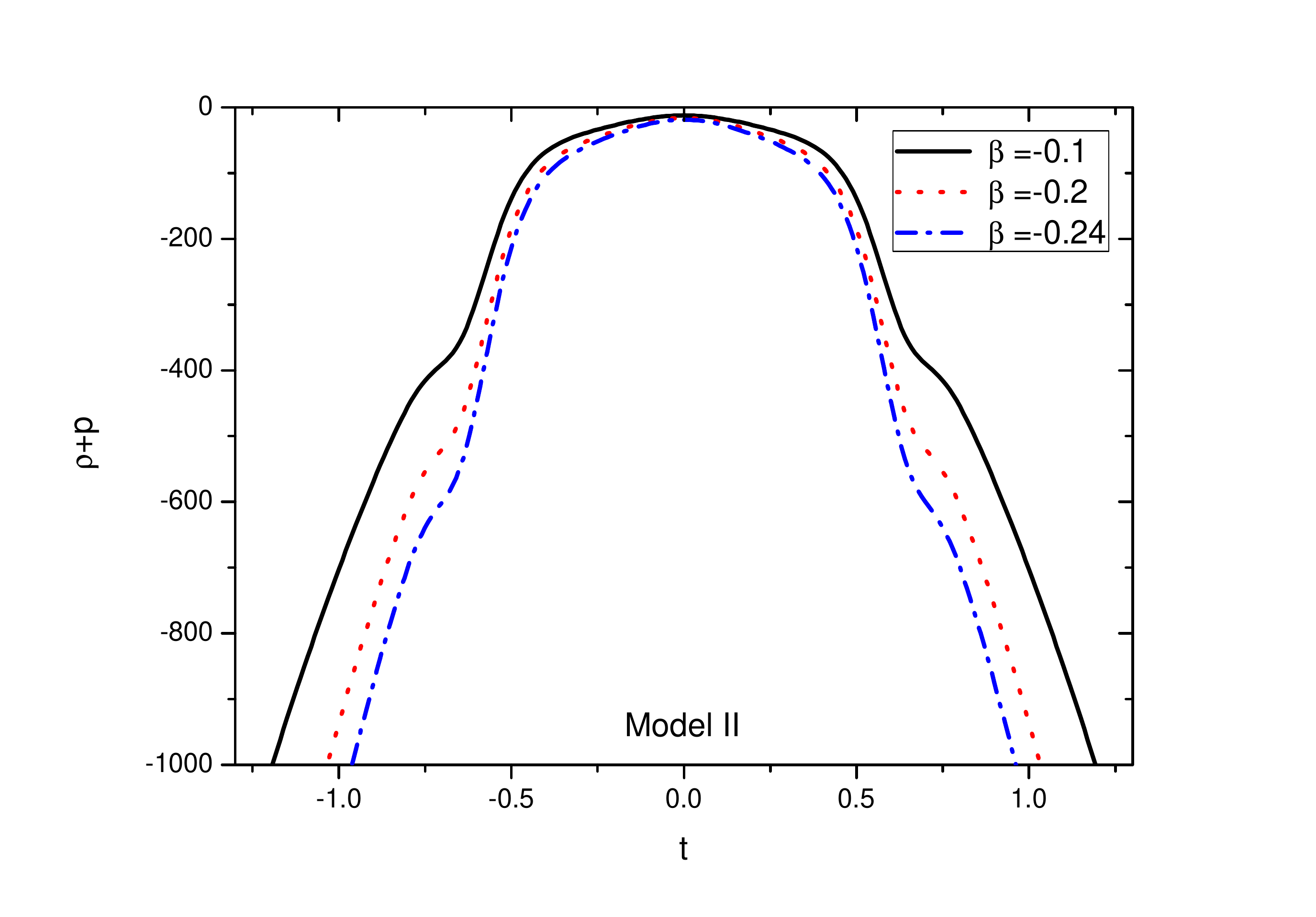}
\caption{Plot for the variation of the $\rho+p$ versus $t$ with the specifications for different parametric values (Model II)}
\end{figure}

\subsection{Model III}
As a third case, we consider a bouncing scale factor given by~\cite{Myrzakulov2014}
\begin{equation}\label{eq:41}
a(t)=a_0e^{(t-t_0)^{2n}},
\end{equation}
where $a_0>0$ is the scale factor at time $t_0$. The exponent $n \neq 0$ decides the bouncing behaviour of the model.

The Hubble parameter for this {\it ansatz} is given by
\begin{equation}\label{eq:42}
H(t)= 2n(t-t_0)^{2n-1},
\end{equation}
so that $\dot{H}=2n(2n-1)(t-t_0)^{2n-2}$.  For $t>0$, the condition $\dot{H}>0$ requires that $n>\frac{1}{2}$. For $n=\frac{1}{2}$, one gets de Sitter type behavior and for $n<\frac{1}{2}$, there is occurrence of future singularities at $t=t_0$. At this singularity the curvature is divergent and separates the expansion and contraction epochs. This model requires $n$ to be positive natural numbers. The model bounces at $t_b=t_0$ when the bouncing scale factor becomes $a_0$. It is obvious that as $t\rightarrow \infty$, we have $a\rightarrow\infty$ and $H\rightarrow \infty$ for positive integral values of $n$.

The deceleration parameter for this bouncing  model becomes
\begin{eqnarray}
q = -1-\frac{2n-1}{2n(t-t_0)^{2n}},\label{eq:43}
\end{eqnarray}

The deceleration parameter is shown in FIG. 1. In the figure we have considered the scale factor at bouncing epoch $a_0=1$, the bouncing epoch $t_0=1$ and the exponent $n=0.6$. The deceleration parameter is a negative quantity for $n > \frac{1}{2}$ and evolves from the bouncing epoch asymptotically to $q=-1$.

\begin{figure}[ht!]
\centering
\includegraphics[width=8cm,height=6cm,angle=0]{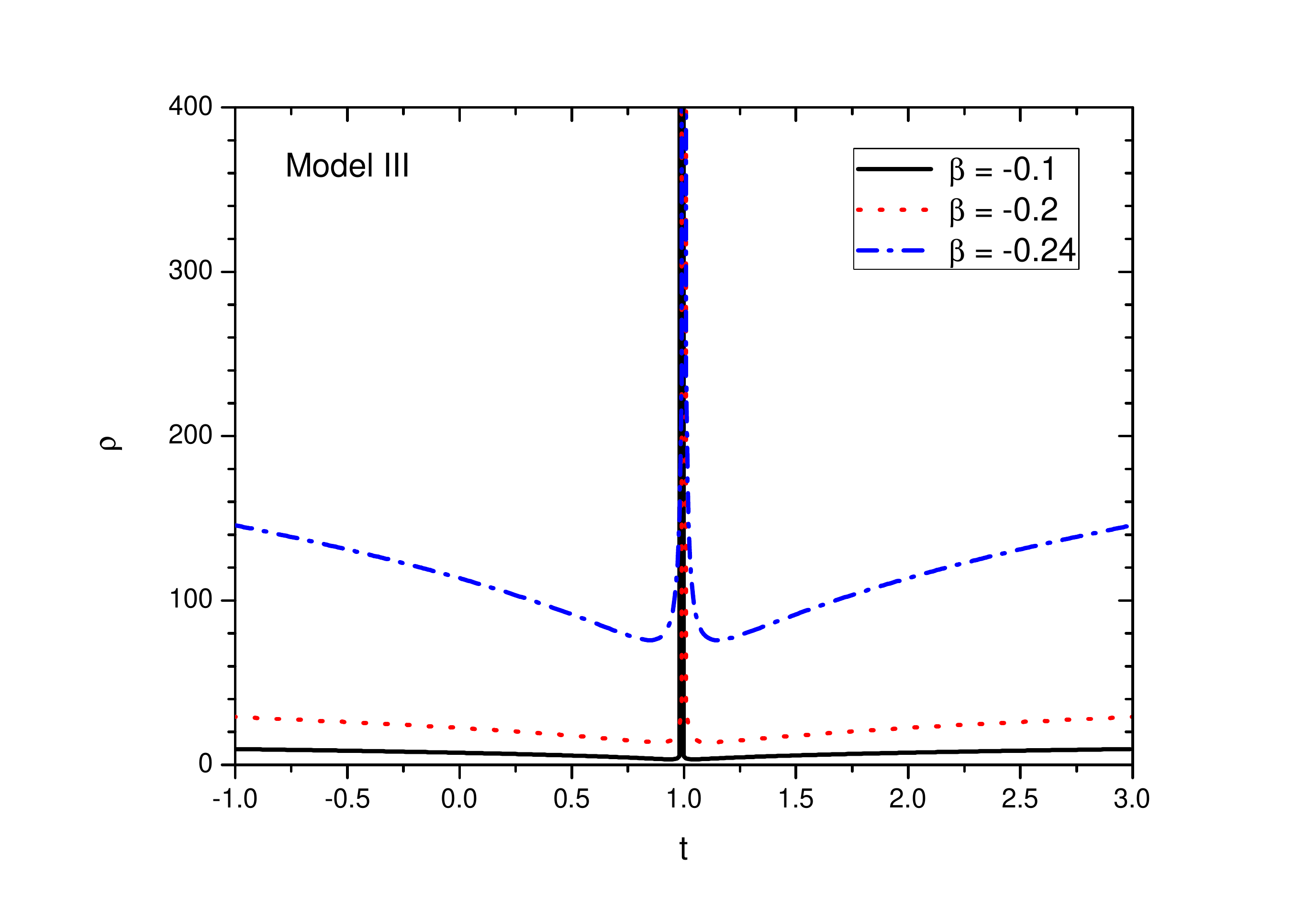}
\caption{Plot for the variation of the $\rho$ versus $t$ with the specifications for different parametric values (Model III)}
\end{figure}

\begin{figure}[ht!]
\centering
\includegraphics[width=8cm,height=6cm,angle=0]{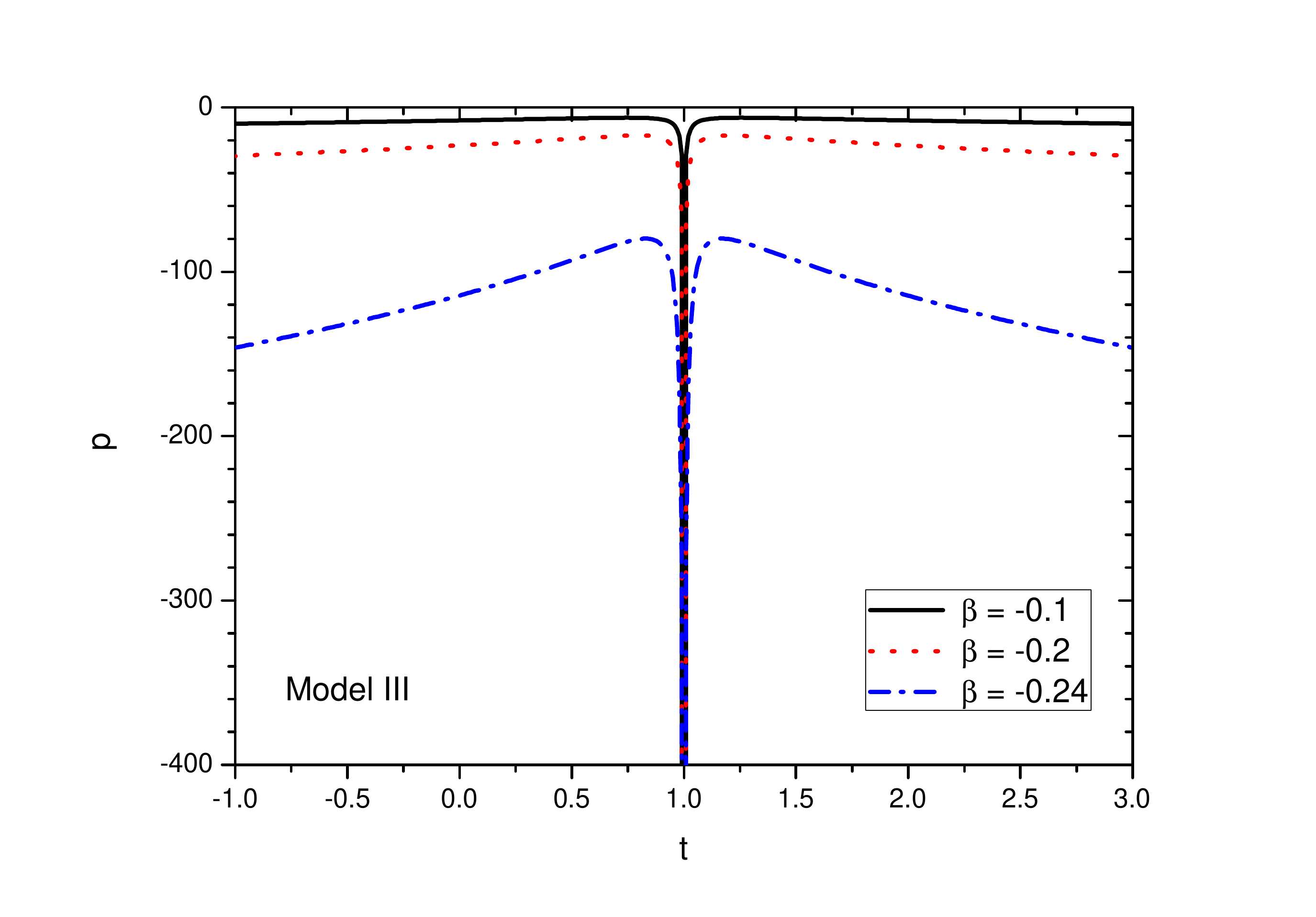}
\caption{Plot for the variation of the $p$ versus $t$ with the specifications for different parametric values (Model III)}
\end{figure}

The evolutionary aspects of the energy density and the pressure for the Model III are shown, respectively, in FIGS. 10 and 11. The parameters are fixed so as to get positive energy density throughout the cosmic evolution and accordingly we obtain a positive energy density in FIG. 10 for all the choices of $\beta$. A sharp peak in the energy density is observed at the bouncing epoch. The pressure is obtained to be negative for all the chosen values of $\beta$. Here also, we note a sharp decrease in pressure near the bounce.

The EOS parameter for this model can be expressed as
\begin{equation}\label{eq:44}
\omega=-1+\frac{\left[2(1+4\beta)\frac{2n-1}{2n(t-t_0)^{2n}}\right]}{2\beta\left[\frac{2n-1}{2n(t-t_0)^{2n}}\right]-3\kappa_T}.
\end{equation}

The EOS parameter evolves asymptotically to overlap with the cosmological constant at late times. At the bouncing epoch, it becomes           $\omega=3+\frac{1}{\beta}$ likewise the previous two Models I and II. In FIG. 12, $\omega$ from Eq. \ref{eq:44} is plotted as a function of the cosmic time for three values of the coupling constant. Here we have used $n=0.6$. The evolutionary aspect of the EOS parameter is similar to that of Model I and Model II. The only difference is that, the sharp well in $\omega$ occurs at the bounce. 

\begin{figure}[ht!]
\centering
\includegraphics[width=8cm,height=6cm,angle=0]{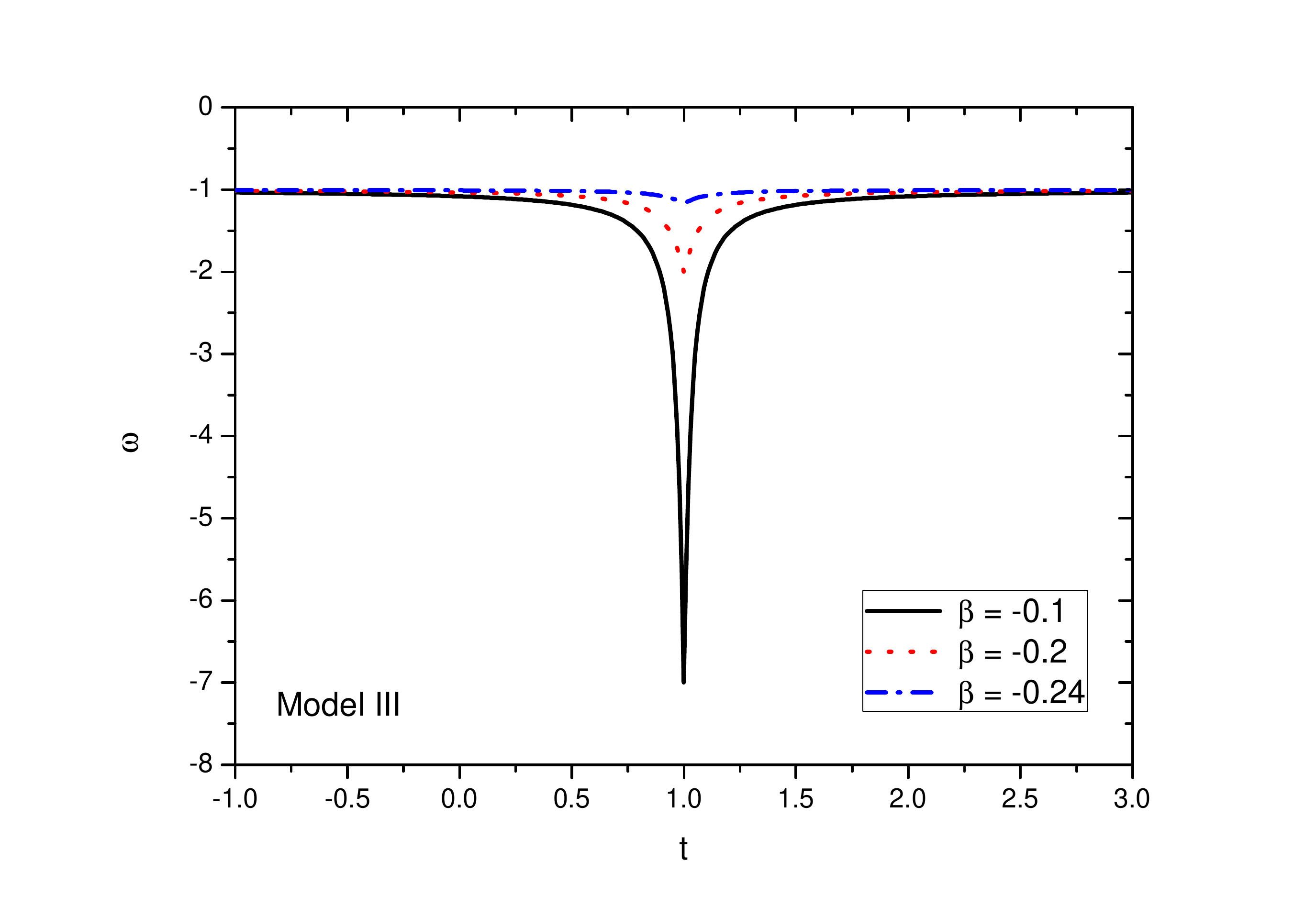} 
\caption{Plot for the variation of the EOS parameter versus time with the specifications of different $\beta$ (Model III)}
\end{figure}

\begin{figure}[ht!]
\centering
\includegraphics[width=8cm,height=6cm,angle=0]{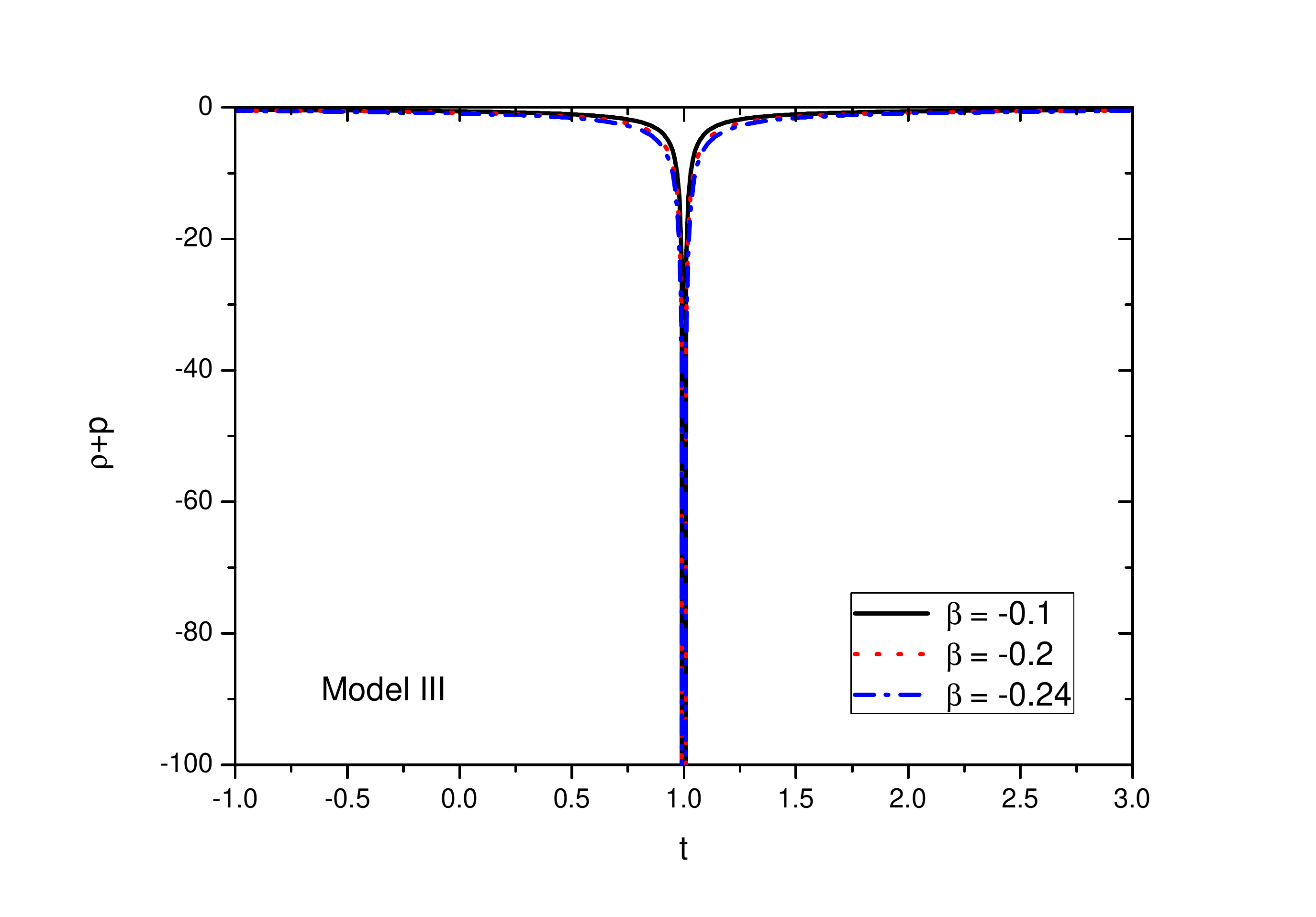}
\caption{Plot for the variation of the $\rho+p$ versus $t$ with the specifications for different parametric values (Model III)}
\end{figure}

The NEC for the present bouncing model is shown in FIG. 13 and as expected, the NEC is violated in the model for the parameter space chosen.

\section{Stability analysis}\label{VIIa}
The stability of the three bouncing  models can be analyzed under linear homogeneous perturbations in the FRW background. We consider linear perturbations for the Hubble parameter and the energy density as~\cite{Sharif2013,Sahoo2020b}
\begin{eqnarray}
H(t)&=& H_b(t)\left(1+\delta(t)\right),\label{eq:45}\\
\rho(t)&=& \rho_b\left(1+\delta_m(t)\right),\label{eq:46}
\end{eqnarray}
where $\delta (t)$ and $\delta_m(t)$ are the perturbation parameters. In the above, we have assumed a general solution $H(t)=H_b(t)$ which satisfies the background FRW equations. The matter energy density can be expressed in terms of $H_b$ as 
\begin{equation}
\rho_b = \frac{3\kappa_T H_b^2-2\beta \dot{H}_b}{\alpha^2-\beta^2}.\label{eq:47}
\end{equation}

In the extended gravity theory with the functional $f(R,T)=R+2\beta T$, the Friedman equation and the trace equation are obtained as
\begin{eqnarray}
\mathbf{\Theta}^2 &=& 3[\rho+2\beta (\rho+p)+f(R,T)],\label{eq:48}\\
R &=& -(\rho-3p)-2\beta (\rho+p)-4f(R,T),\label{eq:49}
\end{eqnarray}
where $\mathbf{\Theta}=3H$ is the expansion scalar. The first order perturbation equation for a standard matter is given by
\begin{equation}
\dot{\delta}_m(t)+3H_b(t)\delta(t)=0.\label{eq:50}
\end{equation}

Using Eqs. \eqref{eq:45} - \eqref{eq:46}, we obtain
\begin{equation}
\alpha T\delta_m(t)=6H_b^2\delta(t).\label{eq:51}
\end{equation}

An elimination of $\delta(t)$ from Eqs. \eqref{eq:50} and \eqref{eq:51} yields the first order matter perturbation equation 
\begin{equation}
\dot{\delta}_m(t)+\frac{\alpha}{2}\frac{T}{H_b}\delta_m(t)=0,\label{eq:52}
\end{equation}
which on integration leads to 
\begin{equation}
\delta_m(t)= C ~exp\left[-\frac{\alpha}{2}\int \frac{T}{H_b}~dt \right],\label{eq:53}
\end{equation}
where $C$ is a non zero positive constant. The evolution of the linear perturbation $\delta(t)$ reads as
\begin{equation}
\delta(t)=\frac{\alpha CT}{6H_b^2}~exp\left[-\frac{\alpha}{2}\int \frac{T}{H_b}~dt \right].\label{eq:54}
\end{equation}

Near the bouncing epoch, we have $H_b(t)=0$ which blows up the factor $\frac{T}{H_b}$ in the integrand of Eqs. \eqref{eq:53} and \eqref{eq:54}. Therefore the model is unstable near the bouncing epoch. However, all the bouncing models considered in the present work, the integral $\int \frac{T}{H_b}~dt$ are evaluated for positive time domain which comes out to be positive. Consequently, the linear perturbation $\delta(t)$ smoothly decay out to provide stability to the models.

\section{Conclusion}
In the present work, we have discussed some bouncing cosmological models in the frame work of a simple extended theory of gravity. The extended theory of gravity is derived from an action where the usual Ricci scalar is replaced by a minimally coupled function linear in $R$ and $T$. The presence of the trace of the energy momentum tensor in the geometry side of the Einstein-Hilbert action leads to an additional interactive term in the field equations. This matter field dependent additional interactive term is responsible for a late time cosmic acceleration as predicted from a lot of observational data. Such a concept to include a bit of matter field within the gravitational action is motivated from quantum effects and particle production process and can be associated with the existence of imperfect fluids. This extended gravity theory is simple to handle mathematically but can be elegant in explaining many issues in cosmology and astrophysics. 

In the frame work of such an extended gravity theory, we have investigated three different models that show bouncing behavior at some epochs. All the three models  explain the late time acceleration. Model II in the limit $\lambda \ll 1$ similar as Model I for $\lambda_1=\lambda_2=\lambda$. However, the deceleration parameter shows contrasting behavior for Model I and Model II although the models share some similarities. In the first two models the bounce takes place at $t_b=0$ but for Model III the bounce takes place at a finite time $t_0$. There might be finite time singularities present in Model III for non-integral positive values of $n$ due to the occurrence of a saddle point in the scale factor at $t=t_0$. In  these models we have shown the effect of the coupling parameter of the extended gravity theory in lifting the omega-singularity occurring in GR. The presence of a finite non-zero but small values of $\lambda$ removes the singularity occurring in the equation of state parameter during the bouncing epoch. 

It is important to note that, the evolutionary behavior of the EOS parameter is mostly decided by the coupling parameter of the extended gravity. We have derived the energy conditions for the models  and observed that for all the models, the energy conditions are violated. In fact, a violation of energy condition is an integral part in the bouncing scenario. Also, the violation of energy conditions enable the model with a positive slope for the Hubble parameter to evolve in the phantom phase. We have analyzed the stability of the bouncing models under a linear homogeneous perturbations in the FRW background. At the bouncing epoch, the models are highly unstable but as we move away from the bounce in the positive time domain, the perturbation smoothly decay out ensuring stability of the models at late times.

\section*{ACKNOWLEDGMENTS}
S.K.T., B.M., and S.R. are thankful to the Inter-University Centre for Astronomy and Astrophysics, Pune, India for providing the Visiting Associateship under which a part of this work was carried out.

\end{document}